\documentstyle[aps,prl]{revtex}
\begin{document}
\draft \twocolumn[\hsize\textwidth\columnwidth\hsize\csname
@twocolumnfalse\endcsname 
\title{ Low-energy singlet and triplet
excitations in the spin-liquid phase of the two-dimensional
$J_{1}-J_{2}$ model }
%\author{Authors}
\author{Valeri N. Kotov$^{(1,2)}$, J. Oitmaa$^{(1)}$, Oleg P. Sushkov$^{(1)}$, and
 Zheng Weihong$^{(1)}$} \address{
$^{(1)}$School of Physics, University of New South
 Wales, Sydney 2052, Australia\\
$^{(2)}$Department of Physics, University of Florida,
Gainesville, FL 32611-8440\\
}

%\date{\today}
\maketitle
\begin{abstract}
We analyze the stability of the spontaneously dimerized 
 phase of the frustrated Heisenberg antiferromagnet - the
 $J_{1}-J_{2}$ model, in two dimensions. The lowest triplet
 excitation, corresponding to breaking of a singlet bond, is found to
 be stable in the region $0.38 \lesssim J_{2}/J_{1} \lesssim 0.62$.  In
 addition we find a stable low-energy collective singlet mode, which
 reflects  the spontaneous violation of the discrete
 symmetry.  The spontaneous dimerization vanishes at the point of
 second-order quantum transition into the N\'{e}el ordered phase
 ($(J_{2}/J_{1})_{c1} \approx 0.38$).  We argue that the disappearance of
 dimer order is related to the vanishing of the singlet energy gap at
 the transition point.

\end{abstract}

 ]

%Introduction.

The nature of excitations in the quantum disordered phases of
 low-dimensional quantum antiferromagnets is a topic of fundamental
 importance for the physics of quantum magnetism \cite{subir}.  Such
 phases can result from mobile holes in an antiferromagnetic
 background as in the $t-J$ or Hubbard model at finite doping.
 Alternatively, competition of purely magnetic interactions can also
 lead to destruction of long-range order.  A typical example of the
 second kind is the $J_{1}-J_{2}$ model which exhibits a quantum
 disordered (spin-liquid) phase due to second-neighbor frustrating
 interactions. Even though it has been intensively studied during the
 last ten years, the $J_{1}-J_{2}$ model still holds many secrets,
 especially concerning its quantum disordered phase.  Exact
 diagonalization studies \cite{exact} have shown that the excitation
 spectrum of the model is quite complex and that finite-size effects
 are large \cite{exact1}. Spin-wave like expansions around the ordered
 phase (which occurs for small frustration) naturally can not give any
 information about the disordered phase at stronger frustration, and
 consequently non-perturbative methods are needed to analyze the
 latter regime.

  A new insight into the disordered regime was achieved by
 field-theory methods \cite{Read,subir}, dimer series expansions
 \cite{series} and related effective Hamiltonian approaches
 \cite{boson,boson1}.  The above works have shown that the ground
 state in the disordered regime is dominated by short-range singlet
 (dimer) formation in a given pattern (see Fig.1.). The stability of
 such a configuration implies that the lattice symmetry is
 spontaneously broken and the ground state is four-fold
 degenerate. Such a route towards quantum disorder is known rigorously
 to take place in one dimension, where the Lieb-Schultz-Mattis (LSM)
 theorem guarantees that a gapped phase always breaks the
 translational symmetry and is doubly degenerate, whereas gapless
 excitations correspond to a unique ground state \cite{LSM}. A
 generalization of the LSM theorem to higher dimensions has been
 suggested \cite{Affleck}, but can not be rigorously proven.

In this Letter we study further the stability of the spontaneously
 dimerized phase in the $J_{1}-J_{2}$ model and find that in the
 disordered phase not only the one-particle triplet excitations but
 also the two-particle singlet modes are stable. Both types of
 excitations are gapped and become gapless at the transition point to
 the N\'{e}el phase. The singlet collective mode is a low-energy 
 excitation of the system and we argue that it is responsible for the
 decrease and ultimately the vanishing of the dimer order parameter as the transition
 is approached from the disordered side. Since the excitation spectrum
 is complex and close to the transition point the triplet is not the
 only low-energy excitation, it is quite possible, in our view, that
 the $O(3)$ non-linear sigma model in its usual formulation
 \cite{subir} is not the correct effective field theory for the
 frustrated antiferromagnet.

%One-particle spectrum.

The Hamiltonian of the $J_{1}-J_{2}$ model reads:

\begin{equation}
\label{ham}
H = J_{1} \sum_{nn}{\bf S}_{i}.{\bf S}_{j} + J_{2} \sum_{nnn} {\bf
  S}_{i}.{\bf S}_{j},
\end{equation}
where $J_{1}$ is the nearest-neighbor, and $J_{2}$ is the frustrating
  next-nearest-neighbor Heisenberg exchange on a square lattice (see
  Fig.1).  Both couplings are antiferromagnetic, i.e.  $J_{1,2}>0$ and
  the spins $S_{i}=1/2$.  In order to study the excitations and their
  stability in the dimer phase we first derive an effective
  Hamiltonian in terms of bosonic  operators creating triplets $t_{i
  \alpha}^{\dag}, \alpha=x,y,z$ from the singlets formed by each pair
  of spins, as shown in Fig.1.  Similar effective theories have been
  derived in Refs.\cite{boson,boson1} and we only present the result:
  $ H = H_{2} + H_{3} + H_{4} + H_{U}, $ 
\begin{eqnarray}
\label{ham2}
H_{2} & = &\sum_{\bf{k}, \alpha} \left\{ A_{\bf{k}}
t_{\bf{k}\alpha}^{\dagger}t_{\bf{k}\alpha} +
\frac{B_{\bf{k}}}{2}\left(t_{\bf{k}\alpha}^{\dagger}
t_{\bf{-k}\alpha}^{\dagger} + \mbox{h.c.}\right) \right \}, \\
\label{ham3}
H_{3}& = & \sum_{1+2=3} \mbox{R}({\bf k_{1}},{\bf k_{2}})
\epsilon_{\alpha\beta\gamma} t_{\bf{k_{1}}\alpha}^{\dagger}
t_{\bf{k_{2}}\beta}^{\dagger} t_{\bf{k_{3}}\gamma} + \mbox{h.c.}, \\
\label{ham4}
H_{4} & = & \sum_{1+2=3+4} \mbox{T}({\bf k_{1}}-{\bf k_{3}})\times
\\
&& (\delta_{\alpha\delta}\delta_{\beta\gamma}-
\delta_{\alpha\beta}\delta_{\gamma\delta})
 t_{\bf{k_{1}}\alpha}^{\dagger}
t_{\bf{k_{2}}\beta}^{\dagger}t_{\bf{k_{3}}\gamma}
t_{\bf{k_{4}}\delta}. \nonumber
\end{eqnarray}
 We also introduce an infinite repulsion on each site ($H_{U}$), in order to
 enforce the kinematic constraint on the Hilbert space $t_{i
 \alpha}^{\dag} t_{i \beta}^{\dag} = 0$.
\begin{equation}
\label{hU} 
H_{U} = U \sum_{i,\alpha \beta} t_{\alpha i}^{\dagger}t_{\beta
i}^{\dagger}t_{\beta i}t_{\alpha i}, \ \ U \rightarrow \infty.
\end{equation}
 The constraint is necessary in order to ensure that the bosonic
 Hamiltonian in terms of the triplet operators corresponds uniquely to
 the original spin Hamiltonian (\ref{ham}) and no unphysical states
 appear in the final result.  The introduction of $H_{U}$ into the
 effective theory and its subsequent diagrammatic treatment follow
 closely our previous work Ref.\cite{us1}.
 
The following definitions are used in Eqs.(\ref{ham2}-\ref{ham4})
\begin{eqnarray}
A_{\bf{k}} & = & J_{1} - \frac{J_{1}}{2}\xi_{k_{x}} + (J_{1} -
J_{2})\xi_{k_{y}} - J_{2}\xi_{k_{x}}\xi_{k_{y}}, \nonumber \\
B_{\bf{k}} & = & - \frac{J_{1}}{2}\xi_{k_{x}} + (J_{1} -
J_{2})\xi_{k_{y}} - J_{2}\xi_{k_{x}}\xi_{k_{y}}.
\end{eqnarray}
where $ \xi_{k} = \cos(k), \ \ \gamma_{k} = \sin(k).  $ The matrix
elements in the quartic and cubic interaction terms are:
\begin{eqnarray}
\mbox{T}({\bf k}) & = & \frac{J_{1}}{4} \xi_{k_{x}} +
 \frac{J_{1}+J_{2}}{2}\xi_{k_{y}} +
 \frac{J_{2}}{2}\xi_{k_{x}}\xi_{k_{y}}, \nonumber \\ \mbox{R}({\bf
 p},{\bf q}) & = & - \frac{J_{1}}{4} \gamma_{p_{x}} - \frac{J_{2}}{2}
 \gamma_{p_{x}}\xi_{p_{y}} - \{p \rightarrow q\}.
\end{eqnarray}
Throughout the paper we work in the Brillouin zone of the dimerized
 lattice.
 
The spectrum is determined by  the 
 normal Green's function 
\cite{us1}:
$
G_{N}({\bf{k}},\omega) = [\omega + \tilde{A}_{\bf{k}}(-\omega)]
[ \{ \omega + \tilde{A}_{\bf{k}}(-\omega) \} \{ \omega -
\tilde{A}_{\bf{k}}(\omega) \} +\tilde{B}_{\bf{k}}^{2}]^{-1}
$
where $\tilde{A}_{\bf{k}}(\omega)= A_{\bf{k}} +
\Sigma_{N}({\bf{k}},\omega)$ and $\tilde{B}_{\bf{k}} = B_{\bf{k}} +
\Sigma_{A}({\bf{k}},\omega)$.  The normal self-energy $\Sigma_{N}$ is
given by the diagrams of Fig.2(a) where the effective scattering
vertex $\Gamma$ has to be found by solving the Bethe-Salpeter
equation shown in the same figure. The first two diagrams in Fig.2(a)
(proportional to $\Gamma$ and T) can be calculated as explained in
Ref.\cite{us1}, while the three-particle contribution (third
diagram in Fig.2(a)) was discussed in more detail in Ref.\cite{H3}.
We adopt the Brueckner approximation which is based on the smallness of the
triplet density 
$n_t=\sum_{\alpha}\langle t^{\dag}_{i\alpha}t_{i\alpha}\rangle \ll 1$
 (Ref. \cite{us1}).
However for the system under consideration   $n_t \approx 0.3$ is not very 
small and additional diagrams have to be summed up to ensure the reliability of
 the results. 
 This mainly  concerns the anomalous self-energy
 $\Sigma_{A}$ shown in Fig.2(b).
One improvement was  discussed in Ref.\cite{us3} and involves 
summation of rainbow diagrams containing  $U$.
This is equivalent to introduction of the term 
$\Lambda \sum_{i\alpha}(t^{\dag}_{i\alpha}t^{\dag}_{i\alpha}+h.c.)$ into the
Hamiltonian  and choosing the Lagrange multiplier $\Lambda$ from
the condition $\langle t^{\dag}_{i\alpha}t^{\dag}_{i\alpha} \rangle=0$.
The second improvement is the summation of rescattering diagrams
  which take into account  
the attraction in the singlet channel (see Fig.2(c) and the
discussion of the singlet bound state below). The two effects put together give
\begin{equation}
\label{rb}
\tilde{B}_{\bf{k}}\approx
 - \frac{2J_{1}}{3}\xi_{k_{x}} +
2J_{1}\xi_{k_{y}} - {{2J_1J_2}\over{2J_1-0.5J_2}}\xi_{k_{x}}\xi_{k_{y}}
+\Lambda.
\end{equation}
The pole of $G_{N}({\bf{k}},\omega)$, which has  to be found self-consistently,
  determines the renormalized one-particle spectrum $\omega({\bf k})$.
 Here we only present the results, obtained by a numerical iterative procedure.
  The dot-dashed line in Fig.3. is the spectrum for the  particular value of
 $J_{2}/J_{1}=0.4$. 
For comparison  we have also performed a dimer
 series expansion up to order 8, and have resummed the dimer series by
 using Pad\'{e} approximants \cite{series1}.
The result for the triplet excitation spectrum 
 is plotted as a solid line in Fig.3. 
 The spectra obtained by the two methods are in excellent agreement.
 One can also see that in some regions of ${\bf k}$ space
  $\omega({\bf k})$ lies in 
 the shaded region in Fig.3, which represents the two-particle
 scattering continuum.
%$E_{c}({\bf k}) = \mbox{min}_{\bf q}[\omega({\bf q} + 
%{\bf k}/2)+ \omega({\bf q} - {\bf k}/2)]$. 
The  excitations can decay in that region which is reflected in the poorer
 convergence of the dimer series (larger error bars on the solid curve).

 To map out the phase diagram of the model we have studied
 the instability of the spectrum as $J_{2}/J_{1}$ varies.
  At the critical value  
$(J_{2}/J_{1})_{c1} = 0.38$ the gap at ${\bf k}=(0,\pi)$ (which
 corresponds to the N\'{e}el ordering wave-vector $(\pi,\pi)$ of the 
original (non-dimerized) lattice) vanishes, signaling a transition to
 the N\'{e}el ordered phase. For larger frustration the gap grows almost
 linearly with  $J_{2}/J_{1}$, as shown in  Fig.4(a). At  $(J_{2}/J_{1})_{c2}=0.62$
 an instability towards a collinear phase takes place and the gap
 at $(0,0)$ ($(\pi,0)$ of the original lattice) goes to zero. 
 The transition at  $(J_{2}/J_{1})_{c2}$
 appears to be of first order and is quite similar to the transition
 which occurs in the Heisenberg ladder with frustration \cite{us3}.
  The number of triplets increases sharply near this transition point
 and the excitation spectrum is a complex mixture of one- and many-triplet
 bound states \cite{us3}. In what follows we will  concentrate on the 
 vicinity of the second-order transition into the  N\'{e}el phase.

%Spectrum of the singlet bound state.

 Let us now investigate the spectrum of  collective excitations - in particular we 
 have studied 
 the two-particle bound state with S=0. This state has the form:
$
|\Psi_{{\bf Q}}\rangle= \sum_{\bf q}\Psi
({\bf q},{\bf Q})t_{\alpha, {\bf Q}/2 + {\bf q}}^{\dagger} 
t_{\alpha, {\bf Q}/2 - {\bf q}}^{\dagger}|0\rangle,
$
 where ${\bf Q}$ and ${\bf q}$ are, respectively,
 the total and relative momenta
 of the two quasiparticles.
The wave function $\Psi$ satisfies the
integral equation ($E^{S}({\bf Q})$ is the bound state energy) \cite{us1}:
\begin{eqnarray}
\label{BS}
\lefteqn{
\left [E^{S}({\bf Q})-\omega_{{\bf Q/2}+{\bf q}}-\omega_{{\bf Q/2}- {\bf q}}
\right ]\Psi({\bf q},{\bf Q})=} \hspace{0.7cm} \nonumber \\
& & \sum_{\bf p} \left\{ -2[\mbox{T}({\bf p}-{\bf q}) +
 \mbox{T}({\bf p}+{\bf q})] + U \right \}
\Psi({\bf p}, {\bf Q}).
\end{eqnarray}
Since  we have to take $U \rightarrow \infty$, the following condition
must be imposed via a Lagrange multiplier:  
$
\sum_{\bf q} \Psi({\bf q}, {\bf Q}) = 0. 
$
 The numerical solution of Eq.(\ref{BS}) is presented in Fig.3 with a long
 dashed line. Notice that the  singlet bound state has a very low energy
 and exists below the two-particle continuum for all wave-vectors.             
 The singlet is gapped  everywhere in the disordered phase, however
 its energy at ${\bf k}= (0,0)$ approaches zero near the transition point
 to the N\'{e}el phase as shown in Fig.4(a).
 We emphasize that the appearance of a  low energy singlet in the spectrum
 is quite unusual and represents a characteristic feature of frustrated  spin 
 systems.

 Next we calculate the dimer order parameters in the two spatial
 directions, defined as (see Fig.1 for notations):
 $D_{x} =  
 \langle {\bf S}_{2}.{\bf S}_{3} \rangle - 
 \langle {\bf S}_{1}.{\bf S}_{2} \rangle$,
  $D_{y} = 
 \langle {\bf S}_{1}.{\bf S}_{5} \rangle
-\langle {\bf S}_{1}.{\bf S}_{2} \rangle$.
 Dimer series expansions to order 9 have been carried out for
 these quantities and the results are presented in Fig.4(b).
 Both order parameters appear to approach zero at the transition point
 to the  N\'{e}el phase, even though the
 error bars are also quite large.  Our result is very different form
 earlier works \cite{series} where no substantial decrease of
 the order parameter was found. We attribute the difference to the longer
 series we have generated. 
  We have also calculated the dimer order parameter by using the
 diagrammatic approach. The results shown in Fig.4(b) suggest that
 the dimer order is overestimated in this way, since it shows no
 appreciable decrease as frustration decreases.
   However we envisage the following  physical mechanism which
 would explain the discrepancy between the two calculations. 
 As discussed earlier, when the  transition point $(J_{2}/J_{1})_{c1}=0.38$ is 
approached, the energy of the singlet state decreases and ultimately goes to
 zero. Since the ground state is also a singlet, a strong mixing
 between the bound state singlet and the ground state must occur
 near the transition, which  effectively  leads to the increase of  quantum fluctuations.
 Technically this effect can be taken into account by exact
 calculation of the diagram shown in Fig.2(b). This diagram represents a  contribution
 to the anomalous Green's function and thus effectively to 
 the strength of quantum fluctuations. 
The exact two-particle
 scattering amplitude $M$ which appears in Fig.2(b)  satisfies
 the equation shown in Fig.2(c). Notice that the bound state equation
 (\ref{BS}) is the pole of $M$, and it is quite easy to prove
 that as the bound state energy decreases, the contribution
 of the diagram in  Fig.2(b) increases substantially.
 The numerical implementation of this procedure is however quite difficult.
The renormalization of $\tilde B_{\bf k}$ given by Eq.(\ref{rb}) is a step 
in this direction, but does not fully take  into account the effect.
 Nevertheless it is clear that the vanishing of the singlet gap
 at the transition is intimately related to the  increased quantum fluctuations
 and therefore the vanishing of the dimer order parameter. 

Finally we present exact diagonalization data on a small
 cluster which provide further  evidence 
 that  the ground state is spontaneously dimerized.
 Since there
 are four ways the system can choose the dimer pattern in Fig.1, 
the ground state
 is expected to be four-fold degenerate. Indeed, three singlet states
 are clearly seen above the ground state (See Table.I) even though they 
have non-zero excitation energy on a finite cluster (an exponentially
 small splitting is expected between them for large cluster size). 
One also expects that upon 
 introduction of explicit dimerization $\delta$ into the system
 (i.e. coupling (1+$\delta$) on the  bond [12], [34], etc. (see Fig.1.) and
  (1-$\delta$)  on the bonds [23],  [15], etc.), the degeneracy of the ground 
 state will
 be lifted and therefore  the  three zero energy singlets, related to the
  spontaneous dimerization, should disappear. On a finite cluster
 the energies of these singlets should grow as $\delta \times(
\mbox{system size})$. Indeed one can see from Table.I that
 the singlet energies increase with  $\delta$, which proves their 
 symmetry breaking  origin.

%Discussion.

In summary, 
 we have explicitly calculated, for the first time,  the lowest triplet
 and  collective singlet 
excitations of the two-dimensional $J_1-J_2$ model.  
 Both branches of the spectrum  were found to be 
   stable and gapped in the quantum disordered
 phase  and become gapless at the boundary with the N\'{e}el ordered
 phase. We emphasize that  
  a low-lying singlet around ${\bf k} = (0,0)$
 is present only in a {\it spontaneously} dimerized system.
 In contrast, for systems with explicit, exchange driven dimerization, 
  the binding energy for such a singlet it strictly zero.
 Thus the low-energy singlet around the Brillouin zone center should
 be, in principle, observable in quasi two-dimensional systems
 exhibiting a quantum critical point, separating a dimerized (spin-phonon
 interaction driven) and a 
 N\'{e}el phase \cite{exp}. In addition, we have 
 found that the dimer order parameter tends to zero at the transition point
 and have argued that the disappearance of spontaneous dimer order
 is tied to the vanishing of the singlet gap.
 Let us mention that the  structure of the spectrum, found in the present work,
 supports  the picture of Read and Sachdev \cite{Read}, based on their
 analysis of the $Sp(N)$ field theory. In particular, both in our and
 in the field-theory approach, a singlet mode associated with the spontaneous
 dimer order is present.

 The example of the $J_{1}-J_{2}$ model provides further support for the 
 idea, put forward some time ago \cite{Affleck}, that the
 Lieb-Schultz-Mattis theorem  can be extended to higher dimensions, and
 the gapped states of quantum systems necessarily break the discrete 
 symmetries of the lattice.
 In this connection  it would be 
 very interesting to investigate if  similar conclusions can be reached
 for systems with finite doping, e.g. the $t-J$ model away from half filling.
 The recent discovery of stripes in the high-$T_{c}$ materials \cite{stripes},
 as well as the analysis of the $Sp(2N)$  $t-J$ model for large $N$ 
\cite{mat},
 strongly suggest that the disordered ground states at finite doping
 may be unstable towards configurations which break the lattice symmetries. 
 Further analysis of these possibilities is highly desirable.  
 
We thank M. Kuchiev,  J. Richter, and S. Sachdev 
 for stimulating discussions
 and S. Sachdev for bringing Ref.\cite{exp} to our attention.
This work was supported by a grant from the Australian Research
 Council. V.N.K.  acknowledges the  financial  support of
 NSF Grant DMR9357474 at the University of Florida.

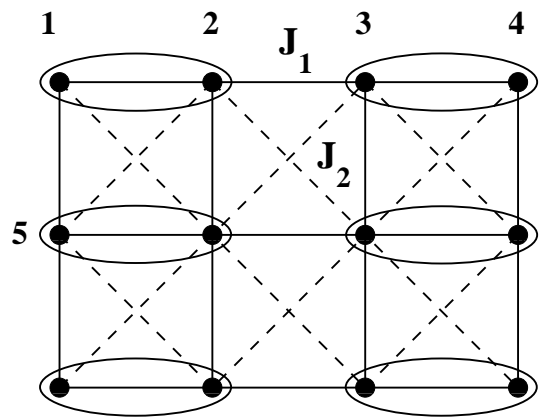
\begin{figure}
\caption
{Schematic representation of the system. The ovals represent
 two spins coupled into a singlet.}
\label{fig.1}
\end{figure}

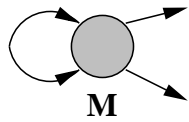
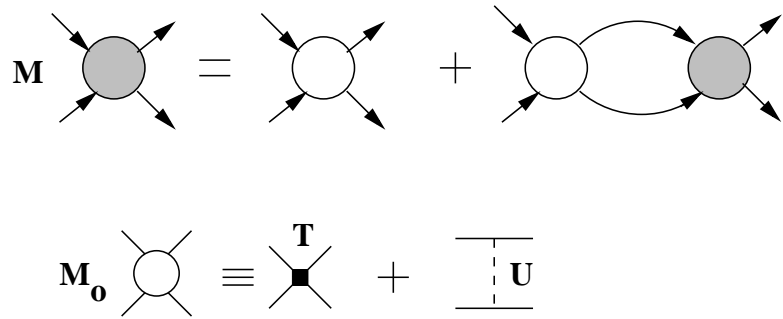
\begin{figure}
\caption
{(a) Lowest order  diagrams for  the normal self-energy $\Sigma_{N}$. 
(b) Contribution to the anomalous self-energy  $\Sigma_{A}$. 
(c) The Bethe-Salpeter equation for the two-particle
 scattering amplitude $M$.}
\label{fig.2}
\end{figure}

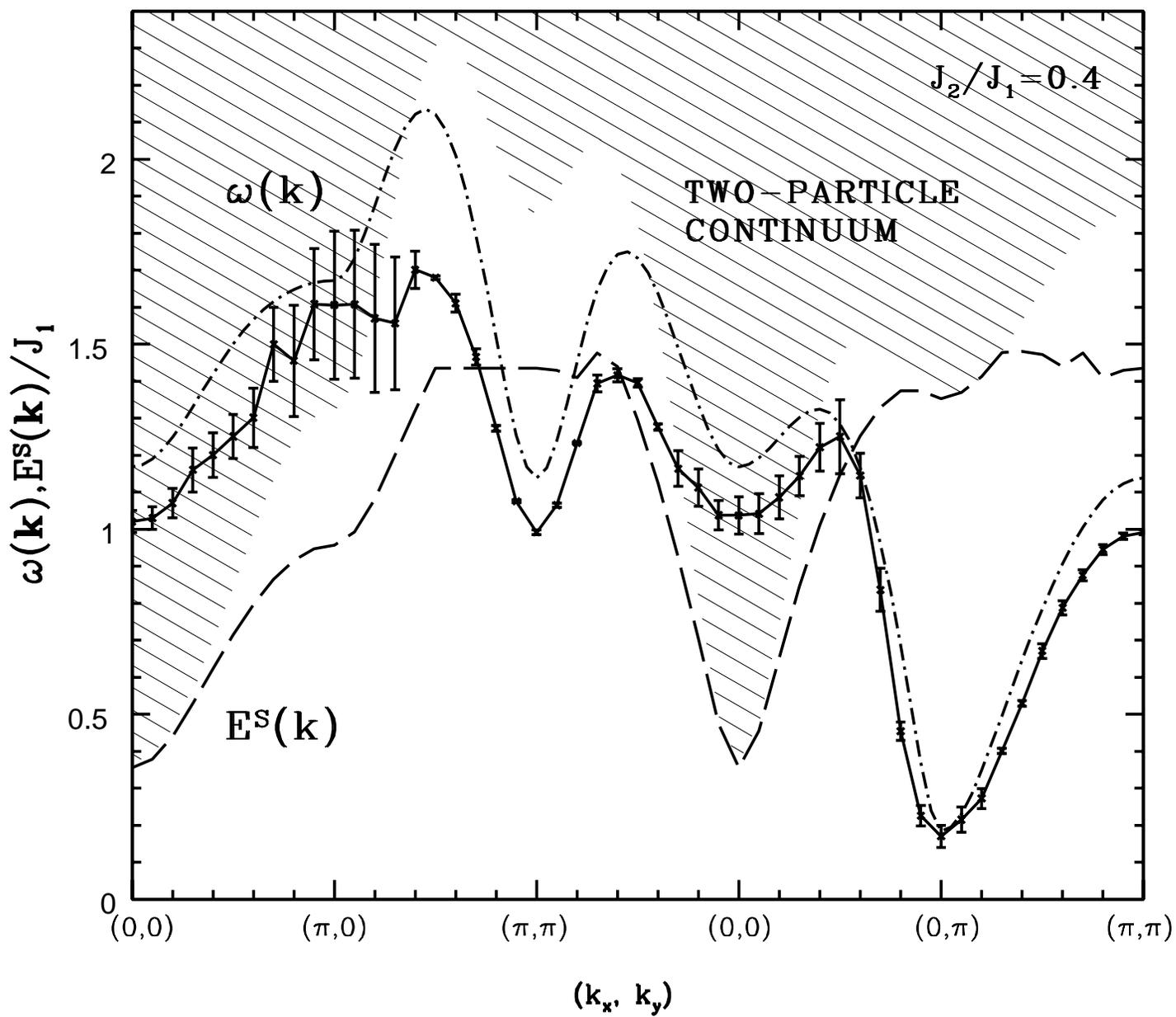
\begin{figure}
\caption
{Excitation spectrum for $J_{2}=0.4J_{1}$. The dot-dashed and solid
 lines are the diagrammatic and dimer series results for the triplet
 dispersion, respectively.
 The long dashed line is the two-particle  singlet excitation.
 The shaded area represents the two-particle scattering continuum.
}
\label{fig.3}
\end{figure}

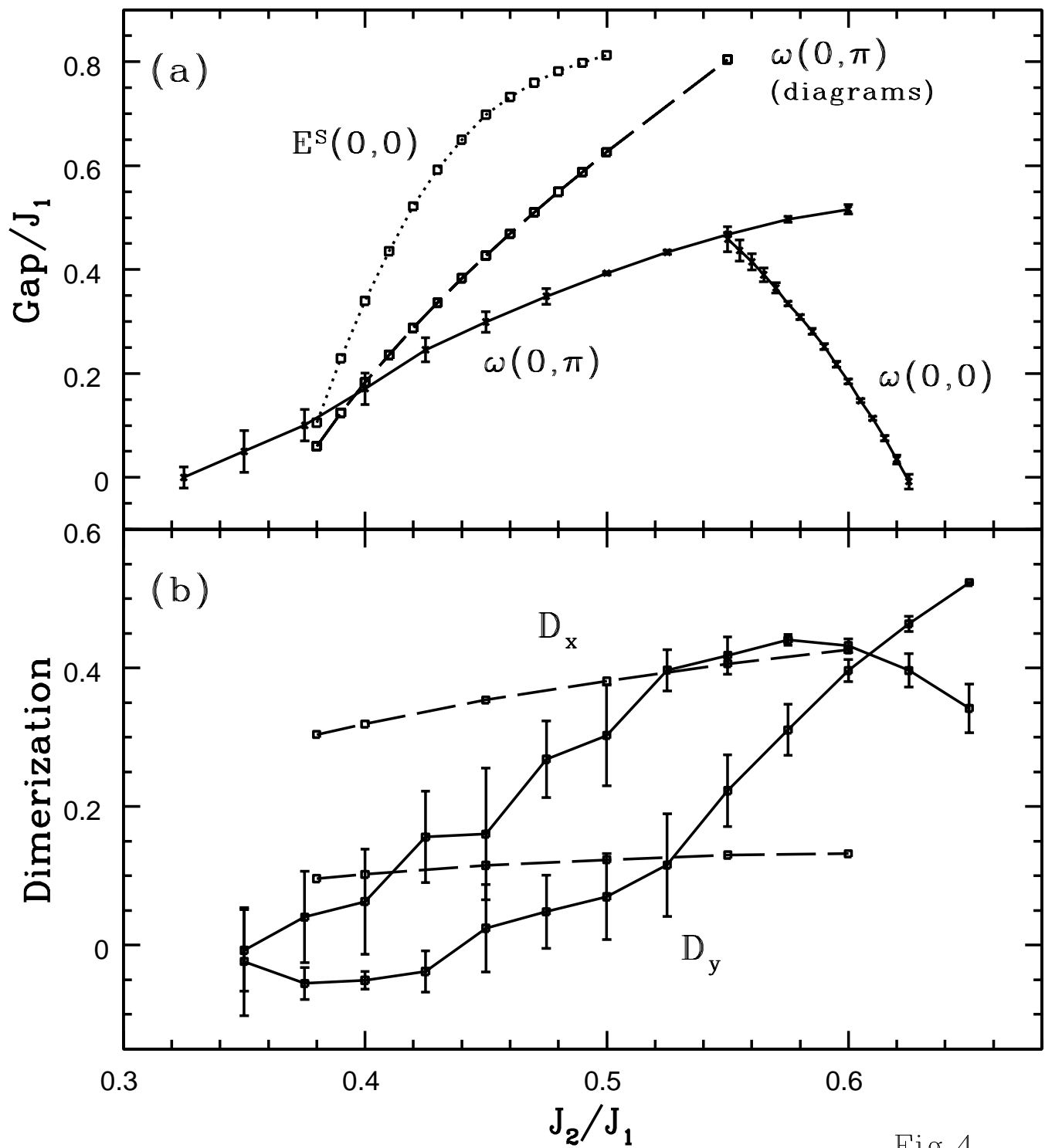
\begin{figure}
\caption
{(a) Triplet and singlet energy gaps in the disordered phase.
 Solid lines represent the dimer series results for the triplet
 gaps. The dotted and  dashed lines are the diagrammatic calculations
 of the singlet and triplet gaps, respectively.
 (b) The dimer order parameters $D_{x}$ and $D_{y}$ (see text for
 definitions). Dashed lines are the diagrammatic results and solid lines are  
 the dimer series results. }
\label{fig.4}
\end{figure}

%\widetext

\begin{table}
\caption{
Exact diagonalization results  on N=$4 \times 4$ cluster 
for the energies of the lowest five excited states 
 at ${\bf k} = (0,0)$.
Frustration is fixed at $J_{2}/J_{1}=0.5$ and $\delta$ is the
 explicit dimerization. "s" and "t" denote   singlet and  triplet levels.  }
\begin{tabular}{lccc}
$\delta=0$ & 0.1 & 0.2 & 0.3 \\
\tableline
0.319 (s) & 0.655 (s) & 0.908 (t) &  0.864 (t)\\
0.830 (s) & 0.864 (s)  & 0.924 (s)& 1.130 (s)\\
0.971 (s)& 1.001 (t) & 1.327 (s) & 1.814 (s)\\
1.128 (t)& 1.144 (s) &1.506 (s) & 2.001 (s)\\
1.135 (s)& 1.425 (t) & 1.684 (t)& 2.028 (t)\\
\end{tabular}
\label{Table 1}
\end{table}

\begin{references}

\bibitem{subir} S. Sachdev, in {\it Low Dimensional
 Quantum Field Theories for Condensed Matter Physicists},
 edited by L. Yu, S. Lundqvist, and G. Morandi, World Scientific, 1995.

\bibitem{exact} E. Dagotto and A. Moreo,  Phys. Rev. Lett. {\bf 63},
 2148 (1989); F. Figueirido {\it et al.}, Phys. Rev. B {\bf 41}, 4619
 (1989).

\bibitem{exact1} H.J. Schulz, T.A.L. Ziman, and D. Poilblanc,
 J. Phys. (Paris) I {\bf 6}, 675 (1996). 

\bibitem{Read} N. Read and S. Sachdev, Phys. Rev. Lett.
{\bf 66}, 1773 (1991); {\it ibid.} {\bf 62}, 1697 (1989).

\bibitem{series}M.P. Gelfand, R.R.P. Singh, and D.A. Huse,
 Phys. Rev. B  {\bf 40}, 10801 (1989); M.P. Gelfand, {\it ibid.}  {\bf 42},
 8206 (1990). 

\bibitem{boson}
 S. Sachdev and R.N. Bhatt,  Phys. Rev. B
 {\bf 41}, 9323 (1990).

\bibitem{boson1}
A.V. Chubukov and T. Jolicoeur, Phys. Rev. B
{\bf 44}, 12050 (1991).

\bibitem{LSM} E. Lieb, T. Schultz, and D. Mattis, Ann. Phys. (N.Y.)
  {\bf 16}, 407 (1961).

\bibitem{Affleck} I. Affleck,  Phys. Rev. B {\bf 37}, 5186 (1988).

\bibitem{us1} V.N. Kotov  {\it et al.}, 
 Phys. Rev. Lett. {\bf 80}, 5790 (1998); 
 O.P. Sushkov and V.N. Kotov,  {\it ibid.} {\bf 81},
1941 (1998).

\bibitem{H3} V.N. Kotov, J. Oitmaa, and W.H. Zheng, 
 Phys. Rev. B {\bf 59}, 11377  (1999).
% P.V. Shevchenko
% {\it et al.},  
% cond-mat/9901302.

\bibitem{us3} V.N. Kotov, O.P. Sushkov, and R. Eder,
 Phys. Rev. B {\bf 59}, 6266 (1999).

\bibitem{series1} For description of linked-cluster series
 expansions see H.X. He {\it et al.}, J. Phys.  A
{\bf  23}, 1775 (1990); M.P. Gelfand {\it et al.}, J. Stat. Phys.
{\bf  59}, 1093 (1990).

\bibitem{exp} D.S. Chow {\it et al.}, Phys. Rev. Lett. {\bf 81},
3984 (1998).


\bibitem{stripes} J.M. Tranquada {\it et al.}, Nature {\bf 375}, 561 (1995).

\bibitem{mat} M. Vojta and S. Sachdev, Phys. Rev. Lett. {\bf 83}, 3916 (1999).

\end{references}
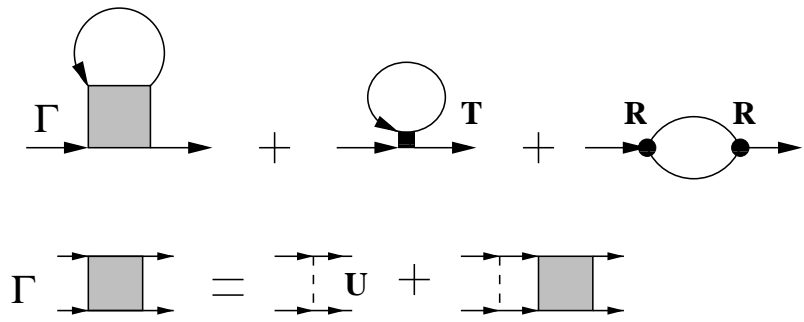
\end{document}